\documentclass[12pt]{article}
\usepackage{epsf,psfrag}
\catcode`\@=11
\usepackage{cite}
\usepackage{xypic}
\usepackage{amsmath}
\usepackage{amsthm}
\usepackage{amsfonts}
\usepackage{amssymb}
\usepackage[cp1251]{inputenc}
\def \be  {\begin{equation}}
\def \ee  {\end{equation}}
\def \ba  {\begin{eqnarray}}
\def \ea  {\end{eqnarray}}
\def\bea{\begin{eqnarray}}
\def\eea{\end{eqnarray}}
\def \bb  {\begin {thebibliography} }
\def \eb  {\end{thebibliography}}
\def \lab #1 {\label{#1}}

\def \matrix #1 {\left(\begin{array}{cc} #1 \end{array}\right)}

\def \Im {\mathop{\rm Im}\nolimits}
\def \Re {\mathop{\rm Re}\nolimits}

\newcommand{\as}{\ifmmode\alpha_{\rm s}\else{$\alpha_{\rm s}$}\fi}
\newcommand{\asbar}{\ifmmode\bar{\alpha}_{\rm s}\else{$\bar{\alpha}_{\rm s}$}\fi}

\font\cmss=cmss12 
\def\inbar{\,\vrule height1.5ex width.4pt depth0pt}
\def\IC{\relax\hbox{$\inbar\kern-.3em{\rm C}$}}
\def\IZ{\relax{\hbox{\cmss Z\kern-.4em Z}}}
\def\IR{{\hbox{{\rm I}\kern-.2em\hbox{\rm R}}}}

\def\IP{{\hbox{{\rm I}\kern-.2em\hbox{\rm P}}}}
\def\II{\hbox{{1}\kern-.25em\hbox{l}}}

\newbox\lett\newdimen\lheight\newdimen\lwidth
\def\ontop#1#2{\setbox\lett=\hbox{#2}\lheight\ht\lett
\multiply\lheight by 12 \divide\lheight by 10\relax%
\lwidth\wd\lett \multiply\lwidth by 8 \divide\lwidth by 10\relax #2\kern-\lwidth%
\raise\lheight\hbox{{$\scriptstyle #1$}}\kern.1ex}

\def\inbar{\,\vrule height1.5ex width.4pt depth0pt}

\numberwithin{equation}{section}
\begin{document}

\begin{titlepage}

\vspace*{1cm}

\begin{center}
{\Large \bf{3j-symbol for the modular double of
$\mathrm{SL}_q(2,\mathbb{R})$ revisited}}

\vspace{1cm}

{\large \sf S.E.Derkachov$ $\footnote{e-mail: derkach@pdmi.ras.ru},
L.D.Faddeev$ $\footnote{e-mail: faddeev@pdmi.ras.ru} \\
}

\vspace{0.5cm}

%\begin{itemize}
%\item[$^a$]
{\it St. Petersburg Department of Steklov Mathematical Institute
of Russian Academy of Sciences,
Fontanka 27, 191023 St. Petersburg, Russia}
%\item[$^b$]
%{\it  Bogoliubov Lab. of Theoretical Physics,
%JINR, 141 980 Dubna, Moscow Reg.,  \\
%and ITPM, M.V.Lomonosov Moscow State University, Russia}
%\item[$^c$]
%{\it Chebyshev Laboratory, St.-Petersburg State University,\\
%14th Line, 29b, Saint-Petersburg, 199178 Russia}
%\end{itemize}
\end{center}

\vspace{1.0cm}

\begin{abstract}

Modular double of quantum group $\mathrm{SL}_q(2,\mathbb{R})$
with $|q|=1$ has a series of selfadjoint irreducible
representations $\pi_s$ parameterized by $s\in \mathbb{R}_+$.
Ponsot and Teschner in~\cite{PT} considered a decomposition of the tensor product $\pi_{s_1}\otimes\pi_{s_2}$ into irreducibles. In our paper we give more detailed derivation and some new proofs.
\end{abstract}

\vspace{2cm}

%\end{titlepage}

{\small \tableofcontents}
\renewcommand{\refname}{References}
\renewcommand{\thefootnote}{\arabic{footnote}}
\setcounter{footnote}{0} \setcounter{equation}{0}

\end{titlepage}

%\large

%\thispagestyle{empty}

\section{Introduction}
%\setstretch{1.5}

Conformal Field Theory is one of the main sources of quantum groups.
The very first example of deformed algebra $G_q$ of
functions on $\mathrm{SL}(2)$ was given by the monodromy matrix
for the quantized Lax operator of the Liouville model~\cite{FT}.
The variable $\tau$, entering the deformation parameter $q = e^{i\pi \tau}$
played the role of the coupling constant.
The duality $\tau \to -\frac{1}{\tau}$ observed in~\cite{G,ZZ} was formalized in~\cite{F} in the notion of modular double.

The irreducible representations of the modular double of $\mathrm{SL}_q(2,\mathbb{R})$, introduced in~\cite{F},
were investigated in~\cite{PT,BT}. In particular in a remarkable paper~\cite{PT} the problem of the decomposition of the tensor product is solved.

There is an intriguing connection of the representations of
the modular double of $\mathrm{SL}_q(2,\mathbb{R})$ and
primary fields of the Liouville model. Both the
representation $\pi_{s}$ of the modular double and vertex operators
$V_{\alpha}(x) = \exp(\alpha\,\phi(x))$ are labeled by the same number
$\alpha = \frac{1}{2}+ i s\,, s>0$. Apparently there should be a correspondence between the operator expansion of $V_{\alpha_1}(x_1)\,V_{\alpha_2}(x_2)$ and the decomposition of $\pi_{s_1}\otimes\pi_{s_2}$ into irreducibles.
Some indications on such connection can be found in~\cite{T}. However the work in this direction is still to be done. Having this in mind we decided to rederive the results of the paper of Ponsot and Teschner~\cite{PT} and supply more details
of derivations and proofs. Our paper is a complete exposition of the talks of the second author (L.D.F) in the early summer of 2012. The first author (S.E.D) joined the company in the late summer and his contribution let to important improvement of the full exposition.

\section{Modular double of $\mathrm{SL}_q(2,\mathbb{R})$}

The algebra has six generators, combined in
two mutually commuting triplets $E\,,F\,,K$ and $\tilde{E}\,,\tilde{F}\,,\tilde{K}$.
The usual relations~\cite{J,D} for $E\,,F\,,K$
\begin{subequations}\label{EFK}
\begin{align}
K\,E = q^2\,E\,K\,,\\[1mm]
K\,F = q^{-2}\,F\,K\,,\\[1mm]
E\,F-F\,E = \frac{K-K^{-1}}{q-q^{-1}}
\end{align}
\end{subequations}
with $q=\mathrm{e}^{i\pi\tau}$ are supplemented by similar relations for $\tilde{E}\,,\tilde{F}\,,\tilde{K}$ with
$\tilde{q}=\mathrm{e}^{\frac{i\pi}{\tau}}$.
Generators $E\,,F\,,K$ and $\tilde{E}\,,\tilde{F}\,,\tilde{K}$ commute.
The coproduct is given by
$$
\Delta(E) = E_{12} = E_1\otimes K_2+I\otimes E_2\,,
$$
$$
\Delta(F) = F_{12} = F_1\otimes I + K_1^{-1}\otimes F_2\,,
$$
$$
\Delta(K) = K_{12} = K_1\otimes K_2
$$
and similarly for $\tilde{E}\,,\tilde{F}$ and $\tilde{K}$.

We shall use the irreducible representations
which are equivalent to the used in~\cite{PT,BT}.
Let $u$ and $v$ realize the Weyl relations
$$
 u\,v = q^2\,v\,u
$$
and act in $L_2(\mathbb{R})$ by explicit formulae
$$
u\,f(x) = \exp\left(-\textstyle\frac{i\pi\, x}{\omega}\right)\, f(x)
\ \ ;\ \
v\,f(x) = f(x+2\omega^{\prime})\,.
$$
Here $\omega$ and $\omega^{\prime}$ are half-periods which substitute
periods $1$ and $\tau$
$$
\tau = \frac{\omega^{\prime}}{\omega} \ \ ;\ \ \omega\,\omega^{\prime}= -\frac{1}{4} \ \ ;\ \ \omega^{\prime\prime} = \omega+\omega^{\prime}\,.
$$
We shall consider the case $\tau >0$, so that $\omega$ and $\omega^{\prime}$ are pure imaginary with positive imaginary part. Operators $u$ and $v$ are unbounded and can be defined on the dense domain $\mathcal{D}$ consisting of the entire functions $f(x)$, rapidly vanishing at infinity along the lines $\Im x = const$.
For instance we can take $f(x)$ in the form
$$
f(x) = \mathrm{e}^{-\alpha\, x^2}\,\mathrm{e}^{\beta\, x}\, P(x)
$$
for positive $\alpha$, arbitrary complex $\beta$ and polynomial $P(x)$.
Operators $u$ and $v$ are nonnegative and essentially selfadjoint.

The representation $\pi_{s}$ is given by formulae
$$
E = \frac{i}{q-q^{-1}}\, e(s)\,,
$$
$$
F = \frac{i}{q-q^{-1}}\, f(s)\,,
$$
where
$$
e(s) = u^{-1}\,\left(q\,v+Z\right) = \left(q^{-1}\,v+Z\right)\,u^{-1}\,,
$$
$$
f(s) = u\,\left(1+q\,Z^{-1}\,v^{-1}\right) =
\left(1+q^{-1}\,Z^{-1}\,v^{-1}\right)\,u\,,
$$
$$
Z = \exp\left(-\textstyle\frac{i\pi\, s}{\omega}\right)\ \,,\ s\in \mathbb{R}\,,
$$
and
$$
K = v
$$
for all $s$.
The representation for the second triple
is given by similar formulae in terms of
$\tilde{u}\,, \tilde{v}$ and $\tilde{Z}$
$$
\tilde{u} = u^{\,\frac{1}{\tau}}\ ,\
\tilde{v} = v^{\,\frac{1}{\tau}}\ ,\ \tilde{Z} = Z^{\,\frac{1}{\tau}}\,,
$$
so that
$$
\tilde{u}\,f(x) = \exp \left(-\textstyle\frac{i\pi\,x}{\omega^{\prime}}\right)
\, f(x)\ \ ;\ \
\tilde{v}\,f(x) = f(x+2\omega)\,,
$$
and $\tilde{Z} = \exp \left(-\textstyle\frac{i\pi\, s}{\omega^{\prime}}\right)$.
Thus $\tilde{u}\,,\tilde{v}$ and $\tilde{Z}$ are obtained from $u\,,v$ and $Z$ by interchange $\omega\rightleftarrows\omega^{\prime}$.

Operators $\tilde{u}\,,\tilde{v}$ have the same domain and are nonnegative and essentially selfadjoint.

Let us note that there is the second regime for $\tau$ which could be called
real form for $\mathrm{SL}_q(2)$. It is the case $|\tau| = 1$ or $\omega^{\prime} = -\bar{\omega}$. In this case involution interchanges pairs $u\,,v$ and $\tilde{u}\,,\tilde{v}$
$$
u^{*} = \tilde{u}\ \,,\ v^{*} = \tilde{v}\,.
$$
This regime has many interesting features. In particular it corresponds to the value of central charge of Liouville model in the interval between $1$ and $25$.
However in this paper we shall consider only regime $\tau > 0$.

\section{Modular quantum dilogarithm}

We shall use this term for the function
$$
\gamma(x) = \exp\left\{\,-\frac{1}{4}\int_{-\infty}^{+\infty}
\frac{\mathrm{d}\,t}{t}\,\frac{e^{i t x}}{\sin(\omega t)\sin(\omega^{\prime} t)}
\,\right\}\,,
$$
where the contour goes above the singularity at $t = 0$. This function has long history, different names, normalizations and applications. The normalization used in this paper is used in~\cite{FKV,V}.

The term modular quantum dilogarithm is used for the function
$$
\Phi(u) = \gamma(x)\ \,,\ u = \exp\left(-\textstyle\frac{i\pi\, x}{\omega}\right)\,.
$$
The adjective "modular"\ is due to the symmetry of $\Phi(u)$
after exchange $\omega\rightleftarrows\omega^{\prime}$
and term "dilogarithm"\  is due to the asymptotic for $\tau\to 0$
$$
\Phi(u)\to \exp \frac{1}{2\pi i \tau}\, \mathrm{Li}_2(-u)\,,
$$
where
$$
\mathrm{Li}_2(u) = \sum_{n=1}^{\infty} \frac{u^n}{n^2}\,,
$$
containing the Euler dilogarithm. Finally "quantum"\ is a
conventional term for $q$-deformation.

The main property is the functional equation
\begin{equation}\label{rec}
\frac{\Phi\left(q\,u\right)}
{\Phi\left(q^{-1}\,u\right)} = \frac{1}{1+u}\,
\end{equation}
and similar one for the shift with $\tilde{q}$.
In terms of function $\gamma(x)$ these equations look as follows
$$
\frac{\gamma(x+\omega^{\prime})}
{\gamma(x-\omega^{\prime})} = 1+e^{-\frac{i\pi}{\omega}\,x} \ \ ;\ \ \frac{\gamma(x+\omega)}
{\gamma(x-\omega)} = 1+e^{-\frac{i\pi}{\omega^{\prime}}\,x}\,.
$$
We shall need also the reflection formula
\begin{equation}\label{reflex}
\gamma(x)\,\gamma(-x) = e^{i\beta}\,e^{i\pi x^2}\ \,, \
\beta = \frac{\pi}{12}\left(\frac{\omega^{\prime}}{\omega}+
\frac{\omega}{\omega^{\prime}}\right)\,
\end{equation}
and formula for the complex conjugation
\begin{equation}\label{complex}
\overline{\gamma(x)} = \frac{1}{\gamma(\overline{x})}\,.
\end{equation}
Asymptotic behaviour is $\gamma(x)\to 1$ for
$\Re(x)\to +\infty$ and reflection formula~(\ref{reflex}) can be used to get asymptotic for $\Re(x)\to -\infty$.

The function $\gamma(x)$ has pole at the point $x= -\omega^{\prime\prime}$ and zero at the point $x= +\omega^{\prime\prime}$. The first terms of the series expansions are~\cite{FKV,V,K1}
\begin{equation}\label{pole}
\gamma(-\omega^{\prime\prime}+z) = -\frac{1}{2\pi i\, c}\,\frac{1}{z} + \ldots\ \ ;\ \
\gamma(\omega^{\prime\prime}+z) = \frac{2\pi i}{c}\, z +\ldots\,,
\end{equation}
Comparison of~(\ref{reflex}) and~(\ref{pole}) gives
$$
c^{2} = e^{-i\beta}\,e^{-i\pi\, \omega^{\prime\prime\,2}}
= i\,e^{2i\beta}\ \ ;\ \ c = e^{i\beta + \frac{i\pi}{4}}\,.
$$
There are main integral identities~\cite{FKV,V}
\begin{equation}\label{FT1eps}
\int_{\mathbb{R}}
\mathrm{d}\,t\, e^{-2\pi i \,t z}\,\frac{1}{\gamma(\omega^{\prime\prime}-i 0-t)} = c\,\gamma(z-\omega^{\prime\prime})\,,
\end{equation}
\begin{equation}\label{FT2eps}
\int_{\mathbb{R}}
\mathrm{d}\,t\, e^{-2\pi i \,t z}\,\frac{\gamma(x-t)}{\gamma(\omega^{\prime\prime}-i 0-t)} = c\,
\frac{\gamma(x)\,\gamma(z-\omega^{\prime\prime})}
{\gamma(x+z)}\,,
\end{equation}
\begin{equation}\label{FT3eps}
\int_{\mathbb{R}}
\mathrm{d}\,t\, e^{-2\pi i \,t z}\,\frac{\gamma(x-t)\,\gamma(y-t)}
{\gamma(\omega^{\prime\prime}-i 0-t)\,
\gamma(x+y+z+\omega^{\prime\prime}-i 0-t)} =
c\,
\frac{\gamma(x)\,\gamma(y)
\,\gamma(z-\omega^{\prime\prime})}
{\gamma(x+z)\,\gamma(y+z)}\,.
\end{equation}

\subsection{Intertwining operator}

Let us use the functional equation to show equivalence of the representations
$\pi_s$ and $\pi_{-s}$. We should find an operator $A(s)$ such that
\begin{subequations}\label{SysA}
\begin{align}
e(s)\,A(s) = A(s)\,e(-s)\,,\\[1mm]
f(s)\,A(s) = A(s)\,f(-s)\,,\\[1mm]
v\,A(s) = A(s)\,v\,.
\end{align}
\end{subequations}
The last equation indicates that $A(s)$ is an operator of
convolution with kernel of the form
$$
A(x,y\,, s) = A(x-y\,, s) = \int_{-\infty}^{+\infty}\mathrm{d}t\,
e^{2\pi i t(x-y)}\, \hat{A}(t)\,.
$$
It is advisable to make Fourier transform. Let $F$ be operator
$$
\left[F f\right](x) = \int_{-\infty}^{+\infty}\mathrm{d}y\,
e^{-2\pi i x y}\, f(y) = \hat{f}(x)\,.
$$
We have $u\,F = F\,v$ and $v\,F = F\,u^{-1}$ and after
conjugation by $F$ we get from~(\ref{SysA})
$$
v^{-1}\,\left(q\,u^{-1}+Z\right)\, \hat{A}(u) =
\hat{A}(u)\, v^{-1}\,\left(q\,u^{-1}+Z^{-1}\right)\,,
$$
$$
v\,\left(1+q\,Z^{-1}\,u\right)\, \hat{A}(u) =
\hat{A}(u)\, v\,\left(1+q\,Z\,u\right)\,.
$$
We move $v^{-1}$ and $v$ to the right and cancel. After that one obtains the equations
$$
\left(q^{-1}\,u^{-1}+Z\right)\, \hat{A}(q^2 u) =
\hat{A}(u)\,\left(q^{-1}\,u^{-1}+Z^{-1}\right)\,,
$$
$$
\left(1+q^{-1}\,Z^{-1}\,u\right)\, \hat{A}(q^{-2} u) =
\hat{A}(u)\,\left(1+q^{-1}\,Z\,u\right)\,,
$$
which are equivalent and differs by the change $u\to q^2 u$
so that really we have only one equation
$$
\left(1+q\,Z\,u\right)\, \hat{A}(q^2 u) =
\hat{A}(u)\,\left(1+q\,Z^{-1}\,u\right)\,.
$$
It is evident that solution of this equation is
$$
\hat{A}(u) = \frac{\Phi(Z u)}{\Phi(Z^{-1} u)}
$$
and finally
\begin{equation}\label{AI}
A(s) = F^{-1}\,\frac{\Phi(Z u)}{\Phi(Z^{-1} u)}\,F.
\end{equation}
The unitarity of the operator $A(s)$ is evident due to~(\ref{complex}).

\section{The main equation for the 3j-symbol}

In this section we shall solve the system of equations for the function $S(x_1,x_2,x_3|s_1,s_2,s_3)$
\begin{subequations}\label{System}
\begin{align}
e_{12}(s_1,s_2)\,S = e^{\prime}_{3}(s_3)\,S\,,\label{S1}\\[1mm]
f_{12}(s_1,s_2)\,S = f^{\prime}_{3}(s_3)\,S\,,\label{S2}\\[1mm]
K_{12}\,S = K^{\prime}_{3}\,S\,,\label{S3}
\end{align}
\end{subequations}
where operators $e_{12}\,,f_{12}\,,K_{12}$ act
on variables $x_1$ and $x_2$
and $e^{\prime}_{3}\,,f^{\prime}_{3}\,,K^{\prime}_{3}$ act
on variable $x_3$. These operators can be obtained from the transposition
$$
u^{\prime} = u \ \,,\ \ v^{\prime} = v^{-1}
$$
and are given by
$$
e^{\prime}_{3} = \left(Z_3+ q\,v_3^{-1}\right)\, u_3^{-1} =
u_3^{-1}\,\left(Z_3+ q^{-1}\,v_3^{-1}\right)\,,
$$
$$
f^{\prime}_{3} = \left(1+q\,Z^{-1}_3\,v_3\right)\, u_{3} =
u_{3}\,\left(1+q^{-1}\,Z^{-1}_3\,v_3\right) \,,
$$
$$
K^{\prime}_{3}=v_3^{-1}\,.
$$
It is already clear that function $S$ realizes the decomposition of the representation $\pi_{s_1}\otimes\pi_{s_2}$ into irreducibles. More will be said in the end of the section.

Following the previous section we shall often use variables $u_i$ instead of $x_i$.
The equation~(\ref{S3})
\begin{equation}\label{S33}
v_1\,v_2\,v_3\, S = S
\end{equation}
allows to exclude $v_2$ from the equation~(\ref{S1}) and $v_2^{-1}$
from the equation~(\ref{S2}) to get the system of equations for
$v_1^{-1}\,S/S$ and $v_3\,S/S$
$$
\left(u_2+q\,Z_{1}^{-1}\,u_1\right)\,v_1^{-1} S +
\left(q\,Z_2^{-1}\,u_2-q^{-1}\,Z_3^{-1}\,u_3\right)\,v_3 S =
\left(u_3-u_1\right)\,S\,,
$$
$$
\left(Z_1\,u_1^{-1}+q\,u_2^{-1}\right)\,v_1^{-1} S +
\left(Z_2\,u_2^{-1} - q^2\,Z_3\,u_3^{-1}\right)\,v_3 S =
q\,u_1^{-1}\,u_3^{-1}\,\left(u_1-u_3\right)\,S\,,
$$
or after diagonalization
$$
\frac{v_1^{-1}\,S}{S} = q\,Z_1\,
\frac{u_1-u_3}{Z_1\,u_3+q\,Z_2\,Z_3\,u_1}
\,\frac{u_2-Z_2\,Z_3\,u_1}
{q\,u_1+Z_1\,u_2}\,,
$$
$$
\frac{v_3\,S}{S} = q\,Z_2\,Z_3\,
\frac{u_1-u_3}{Z_1\,u_3+q\,Z_2\,Z_3\,u_1}
\,\frac{q\,u_2+ Z_1\,u_3}
{Z_2\,u_3-q\,Z_3\,u_2}\,.
$$
More explicitly these equations read
\begin{equation}
S(q^2\,u_1,u_2,u_3) =
\frac{1}{Z_2\,Z_3}\,
\frac{1-\frac{u_3}{u_1}}{1+\frac{Z_1}{q\,Z_2\,Z_3}\frac{u_3}{u_1}}
\,\frac{1-Z_2\,Z_3\,\frac{u_1}{u_2}}
{1+\frac{q}{Z_1}\frac{u_1}{u_2}}\, S(u_1,u_2,u_3)\,,
\end{equation}
\begin{equation}
S(u_1,u_2,q^{-2}\,u_3) =
\frac{Z_1}{Z_2}\,
\frac{1-\frac{u_3}{u_1}}{1+\frac{Z_1}{q\,Z_2\,Z_3}\frac{u_3}{u_1}}
\,\frac{1+\frac{q}{Z_1}\,\frac{u_2}{u_3}}
{1-\frac{q\,Z_3}{Z_2}\frac{u_2}{u_3}}\, S(u_1,u_2,u_3)\,,
\end{equation}
and can be supplemented by
\begin{equation}
S\left(u_1,q^{-2}\,u_2,u_3\right) =
\frac{1}{Z_1\,Z_3}\,\frac{1-Z_2\,Z_3\,\frac{u_1}{u_2}}
{1+\frac{q}{Z_1}\frac{u_1}{u_2}}\,
\frac{1-\frac{Z_3}{q\,Z_2}\frac{u_2}{u_3}}
{1+\frac{1}{q\,Z_1}\frac{u_2}{u_3}}
\, S(u_1,u_2,u_3)\,,
\end{equation}
which is a corollary of~(\ref{S33}).
Let us look for solution in the form
\begin{equation}\label{subs}
S =
\frac{\Phi\left(\alpha_1\frac{u_2}{u_3}\right)
\,\Phi\left(\alpha_2\frac{u_3}{u_1}\right)
\,\Phi\left(\alpha_3\frac{u_1}{u_2}\right)}
{\Phi\left(\beta_1\frac{u_2}{u_3}\right)
\,\Phi\left(\beta_2\frac{u_3}{u_1}\right)
\,\Phi\left(\beta_3\frac{u_1}{u_2}\right)}
\, S_1
\end{equation}
and due to~(\ref{rec}) the choice
$$
\alpha_1 = -q\,Z_3\,Z^{-1}_2\ \,,\ \
\alpha_2 = -q\ \,,\ \
\alpha_3 = Z^{-1}_1\,,
$$
$$
\beta_1 = Z^{-1}_1\ \,,\ \
\beta_2 = Z_1\,Z^{-1}_2\,Z^{-1}_3\ \,,\ \
\beta_3 = - Z_2\,Z_3\,q^{-1}\,,
$$
reduces the equations to
\begin{equation}
S_1(q^2\,u_1,u_2,u_3) = Z^{-1}_2\,Z^{-1}_3\, S_1(u_1,u_2,u_3)\,,
\end{equation}
\begin{equation}
S_1(u_1,u_2,q^{-2}\,u_3) = Z_1\,Z^{-1}_2\, S_1(u_1,u_2,u_3)\,,
\end{equation}
\begin{equation}
S_1\left(u_1,q^{-2}\,u_2,u_3\right) =
Z^{-1}_1\,Z^{-1}_3\, S_1(u_1,u_2,u_3)\,.
\end{equation}
The dual equations, corresponding to the interchange
$\omega\rightleftarrows\omega^{\prime}$, have the
same solution~(\ref{subs}) and after that the Ansatz
$$
S_1(x_1,x_2,x_3) =
\exp -2\pi i \left(s_1\, x_{23}+s_2\, x_{31}+s_3\, x_{21}\right)\, S_0(x_1,x_2,x_3)\,,
$$
where $x_{ik} = x_i-x_k$, reduce the freedom
to double periodic function $S_0$,
which has to be constant.

Thus the solution of the equations~(\ref{System}) is given by
\begin{equation}\label{sol}
S(x_1,x_2,x_3) =
S_0\, \exp -2\pi i \left(s_1\, x_{23}+s_2\, x_{31}+s_3\, x_{21}\right)
\,\cdot
\end{equation}
$$
\cdot\,
\frac{\gamma(x_{12}-s_1)}{\gamma(x_{12}+s_2+s_3+\omega^{\prime\prime})}\,
\frac{\gamma(x_{23}+s_3-s_2-\omega^{\prime\prime})}
{\gamma(x_{23}-s_1)}\,
\frac{\gamma(x_{31}-\omega^{\prime\prime})}
{\gamma(x_{31}+s_1-s_2-s_3)}\,.
$$
The appearance of $\omega^{\prime\prime}$ is due to sign factors in~(\ref{subs})
$$
-q\,u= \exp -\textstyle\frac{i\pi\, (x-\omega^{\prime\prime})}{\omega}
$$
and the singularities here have to be understood as
$$
\omega^{\prime\prime} \to \omega^{\prime\prime} - i\,0\,,
$$
which will be explained in the course of proof of completeness.
The expression for $S$, equivalent to~(\ref{sol}),
was given in~\cite{PT} without derivation.

Now we can interpret the result in more details. The solution exists for any triplet of real $s_1\,,s_2\,,s_3$ and is unique up to normalization constant.
This means that the representation with "spin"\ $s_3$ enter the tensor product
$\pi_{s_1}\otimes\pi_{s_2}$ once for any $s_3$. This can be formalized by the relation
\begin{equation}\label{exp}
\pi_{s_1}\otimes \pi_{s_2} = \int_{0}^{+\infty}
\mathrm{d}s_3\, \rho(s_3)\, \pi_{s_3}\,.
\end{equation}
We can consider $S\left(x_1\,,x_2\,,x_3\right)$ as the kernel of the integral operator $\mathrm{S}$
$$
\pi_{s_3} \xrightarrow{\mathrm{S}} \pi_{s_1}\otimes \pi_{s_2}\,,
$$
$$
f(x_3) \mapsto
\left[\mathrm{S}\,f\right](x_1,x_2) =
\int_{\mathbb{R}} \mathrm{d} x_3 \,
S\left(x_1\,,x_2\,,x_3\right)\, f(x_3)\,,
$$
and equations~(\ref{System}) in operator form are
\begin{equation}\label{System1}
e_{12}\,\mathrm{S} = \mathrm{S}\,e_{3}\ \,,\
f_{12}\,\mathrm{S} = \mathrm{S}\,f_{3}\ \,,\
K_{12}\,\mathrm{S} = \mathrm{S}\,K_{3}\,.
\end{equation}
The complex conjugate function
$\overline{S\left(x_1\,,x_2\,,x_3\right)}$ has interpretation as
the kernel of the projection operator
$$
\pi_{s_1}\otimes \pi_{s_2} \xrightarrow{\mathrm{P}} \pi_{s_3}\,,
$$
$$
f(x_1,x_2) \xrightarrow{\mathrm{P}}
\left[\mathrm{P}\,f\right](x) =
\int_{\mathbb{R}^2} \mathrm{d}\,x_1\mathrm{d}\,x_2\,
\overline{S\left(x_1\,,x_2\,,x\right)}\, f(x_1,x_2)\,.
$$
The measure $\rho(s)$ in~(\ref{exp}) should be found from normalization condition for the kernel $S(x_1,x_2,x_3)$, which will be obtained in the last section.

\section{Undressing of the Casimir}

To get the normalization for $S$ it is useful to
interpret it as an eigenfunction of the Casimir operator
$$
C_{12} = f_{12}\,e_{12} - q\,K_{12} - q^{-1}\,K^{-1}_{12}\,.
$$
It is clear from~(\ref{System1}), that as a function of $x_1$ and $x_2$, $S$ satisfies the equation
$$
C_{12}\,S = \left( Z_3+Z_3^{-1}\right)\,S\,,
$$
where $s_3$ and $x_3$ play the role of parameters labeling eigenvalue and multiplicity.

Explicitly $C_{12}$ can be written as
$$
C_{12} = Z_2\,\frac{u_1}{u_2} + Z^{-1}_2\,\frac{u_2}{u_1} +
\left(Z_1+\frac{1}{q}\,\frac{u_2}{u_1}\right)
\left(1+\frac{q}{Z_1}\,\frac{u_1}{u_2}\right)\,v_2 +\left(Z_2+\frac{Z_1}{q\,Z_2}\,\frac{u_2}{u_1}\right)
\left(1+\frac{q}{Z_1}\,\frac{u_1}{u_2}\right)
v_1^{-1} +
$$
$$
+ \frac{Z_1}{q^2}\,
\frac{u_2}{u_1}\,\left(1+\frac{q}{Z_1}\,\frac{u_1}{u_2}\right)
\left(1+\frac{q^3}{Z_1}\,\frac{u_1}{u_2}\right)\,v_1^{-1}\,v_2\,,
$$
or
$$
C_{12} = Z_2\,\frac{u_1}{u_2} + Z^{-1}_2\,\frac{u_2}{u_1} +
\left(Z_1+\frac{1}{q}\,\frac{u_2}{u_1}\right)\,V_2 +\left(Z_2+\frac{Z_1}{q\,Z_2}\,\frac{u_2}{u_1}\right)\,V_1^{-1} +
\frac{Z_1}{q^2}\,
\frac{u_2}{u_1}\,V_1^{-1}\,V_2\,,
$$
offer the substitution
$$
V_1 = v_1\,\frac{1}{1+\frac{q}{Z_1}\,\frac{u_1}{u_2}}\ \ ;\ \
V_2 = \left(1+\frac{q}{Z_1}\,\frac{u_1}{u_2}\right)\,v_2\,.
$$
We shall introduce a series of adjoint transformations of $C_{12}$ to reduce it to more simple form. These transformations we shall call "undressing".

The first step is to use operator $R_1$ to cancel factors in front of $v_1$ and $v_2$
$$
R_1^{-1}\,R_1\,S = v_1 \ \ ;\ \ R_1^{-1}\,V_2\,R_1 = v_2\,.
$$
The solution is a multiplication operator by function
$$
R_1 = \Phi\left(\frac{1}{Z_1}\frac{u_1}{u_2}\right)\,.
$$
The operator $K_{12}$ is invariant under transformation $R_1$
$$
R_1^{-1}\,K_{12}\,R_1 = K_{12}
$$
and $C_{12}$ transforms into
$$
C_{12}^{\prime} = R_1^{-1}\,C_{12}\,R_1 =  Z_2\,\frac{u_1}{u_2} +
Z_1\,v_2 +Z_2\,v_1^{-1} + Z^{-1}_2\,u_1^{-1}\,
\left(1+q^{-1}\,Z_1\,v_1^{-1}\right)\,U_2\,,
$$
where
$$
U_2 = u_2\,\left(1+q^{-1}\,Z_2\,v_2\right)\,.
$$
Now we find $R_2$ transforming $U_2$ to $u_2$
$$
R_2^{-1}\,U_2\,R_2 = u_2\,.
$$
It is clear that $R_2$ is similar to $R_1$ after
interchange $u_2$ and $v_2$, which is given by the
Fourier transformation which respect variable $x_2$, so that
$$
R_2 = F_2^{-1}\,\tilde{R}_2\,F_2\,,
$$
where $\tilde{R}_2$ is a multiplication by
$\Phi(Z_2 u_2)$.
The operator $K_{12}$ is invariant under transformation $R_2$ and
operator $C_{12}^{\prime}$ acquires the form
$$
C_{12}^{\prime\prime} = R_2^{-1}\,C_{12}^{\prime}\,R_2
= Z_2\,\frac{u_1}{u_2} + \frac{1}{Z_2}\,\frac{u_2}{u_1} + Z_1\,V^{\prime}_2 +
\frac{Z_1}{q\,Z_2}\,\frac{u_2}{u_1}\,V_1^{\prime\, -1}\,,
$$
where
$$
V^{\prime}_1 = v_1\,\frac{1}{1+\frac{q\,Z_2^2}{Z_1}\,\frac{u_1}{u_2}}\ \ ;\ \
V^{\prime}_2 = \left(1+\frac{q\,Z_2^2}{Z_1}\,\frac{u_1}{u_2}\right)\,v_2\,.
$$
Now we transform $V^{\prime}_1$ and $V^{\prime}_2$ to $v_1$ and $v_2$ by multiplication operator
$$
R_3 = \Phi\left(\frac{Z_2^2}{Z_1}\frac{u_1}{u_2}\right)\,,
$$
which leaves $K_{12}$ invariant and transform $C_{12}^{\prime\prime}$ into
$$
\tilde{C}_{12} = R_3^{-1}\,C^{\prime\prime}_{12}\,R_3 = Z_2\,\frac{u_1}{u_2} + \frac{1}{Z_2}\,\frac{u_2}{u_1} + Z_1\,v_2 +
\frac{Z_1}{q\,Z_2}\,\frac{u_2}{u_1}\,v_1^{-1}\,.
$$
Altogether the operator
\begin{equation}\label{A}
\mathrm{A} = R_1\,F_2^{-1}\,\tilde{R}_2\,F_2\,R_3
\end{equation}
gives $\tilde{C}_{12}$ from $C_{12}$ and leaves $K_{12}$ invariant
$$
\tilde{C}_{12} = \mathrm{A}^{-1}\,C_{12}\,\mathrm{A}\ \ ;\ \
\mathrm{A}^{-1}\,K_{12}\,\mathrm{A} = K_{12}\,.
$$
In more explicit form $\mathrm{A}^{-1}$ act as follows
on the function of two variables $f(x_1,x_2)$
\begin{equation}\label{Akern}
\left[\mathrm{A}^{-1}\,f\right](x_1,x_2) =
\frac{1}
{c\,\gamma(x_{12}-s_1+2\,s_2)}\,\int_{\mathbb{R}}
\mathrm{d}\,t\,
e^{2\pi i \,t (s_2-\omega^{\prime\prime})}\,
\frac{\gamma(-t-\omega^{\prime\prime}+i\,0)}
{\gamma(x_{1}-t-s_1)}\,f(x_1,t)\,.
\end{equation}
Operator $\tilde{C}_{12}$ is much more simple then $C_{12}$ and
the problem of simultaneous diagonalization of
$\tilde{C}_{12}$ and $K_{12}$ allows separation of variables.
Consider equations for the corresponding eigenfunctions
$$
K_{12}\,\Psi_p(x_1,x_2) = v_1\,v_2\,\Psi_p(x_1,x_2) = e^{\frac{i \pi  p}{\omega}}\,\Psi_p(x_1,x_2)
$$
$$
\left(Z_2\,\frac{u_1}{u_2} + \frac{1}{Z_2}\,\frac{u_2}{u_1} + Z_1\,v_2 +
\frac{Z_1}{q\,Z_2}\,\frac{u_2}{u_1}\,v_1^{-1}\right)\,\Psi_p(x_1,x_2) =
\left(Z_3+Z_3^{-1}\right)\,\Psi_p(x_1,x_2)
\,,
$$
where we parameterize the eigenvalues by $p$ and $s_3$.
The first equation allows to exclude $v_{1}^{-1}$ from the second to get
$$
\left[Z_2\,\frac{u_1}{u_2} + \frac{1}{Z_2}\,\frac{u_2}{u_1} +
Z_1\left(1+\frac{e^{-\frac{i \pi  p}{\omega}}}{q\,Z_2}\,\frac{u_2}{u_1}\right)\,v_2\right]\,\Psi_p(x_1,x_2) =
\left(Z_3+Z_3^{-1}\right)\,\Psi_p(x_1,x_2)\,.
$$
The general solution of the first equation is given by
$$
\Psi_p(x_1,x_2) = e^{-2\pi i p\, x_1}\Psi_p(x_{21})\,,
$$
where $x_{21} = x_2-x_1$ and after substitution
$$
\Psi_p(x_{21}) = \frac{e^{-2\pi i s_1\,x_{21}}}{\gamma(x_{21}+p-s_2)}\,
\Psi(x_{21})\,,
$$
which eliminates the factor in front of $v_2$, we get
$$
\left(Z_2\,\frac{u_1}{u_2} + \frac{1}{Z_2}\,\frac{u_2}{u_1} +v_2\right)\,\Psi(x_{21}) =
\left(Z_3+Z_3^{-1}\right)\,\Psi(x_{21})\,.
$$
Introducing new operators
$$
u = \frac{1}{Z_2}\frac{u_2}{u_1}\ \ ;\ \ v=v_2
$$
we rewrite the remaining equation in the form
$$
\left(v+ u +u^{-1}\right)\,\Psi =
\left(Z_3+Z_3^{-1}\right)\,\Psi\,.
$$
The operator in LHS is well known in CFT.
It appears as a trace of monodromy of Lax operator
in the Liouville model~\cite{FT}. In quantum Teichm\"{u}ller thery it
got the name of the length operator for geodesics. R.Kashaev
has shown~\cite{K} that this operator has continues spectrum in the interval $[2,\infty]$ with eigenvalues parameterized in the form $Z+Z^{-1}$ with
$$
Z = \exp\left(-\textstyle\frac{i\pi\, s}{\omega}\right)\ \,, \ s\geq 0
$$
and eigenfunctions given by
\begin{equation}\label{phi}
\phi(x,s) = e^{-i\pi(x-\omega^{\prime\prime})^2}\,
\gamma(x+s-\omega^{\prime\prime}+i\,0)\,
\gamma(x-s-\omega^{\prime\prime}+i\,0)
\end{equation}
an even functions of $s$, so that they can be considered for any
$s\in \mathbb{R}$. Kashaev proved~\cite{K} the orthogonality and completeness
for $\phi(x,s)$ in the form
\begin{equation}\label{ort}
\int_{\mathbb{R}} \mathrm{d}\,x\,
\overline{\phi(x,s)}\,\phi(x,s^{\prime}) = \rho^{-1}(s)\,
\left[\,\delta(s-s^{\prime})+\delta(s+s^{\prime})\,\right]\,,
\end{equation}
\begin{equation}\label{compl}
\int_{0}^{+\infty} \mathrm{d}s\,\rho(s)\,
\overline{\phi(x,s)}\,\phi(y,s) = \delta(x-y)\,,
\end{equation}
with $\rho(s)$ given by
$$
\rho(s) = M(s)\,M(-s) = -4\sin\frac{\pi s}{\omega}\,
\sin\frac{\pi s}{\omega^{\prime}}\,,
$$
where $M(s)$, which can be considered as analogue of the Jost function from scattering theory or Harish-Chandra-Gindikin-Karpelevich function from the theory of representations of $\mathrm{SL}(2,\mathbb{R})$, can be taken as
$$
M(s) = c\,e^{-2i\pi s^2 - 2i\pi s\omega^{\prime\prime}}\,
\gamma(2s+\omega^{\prime\prime})\,.
$$
One can say, that the operator
$$
\left[U\,f\right](s) = \int_{-\infty}^{+\infty}\mathrm{d} x
\,M(s)\,\phi(x,s)\, f(x) = F(s)
$$
acts from $L_2(\mathbb{R})$ into the subspace of $L_2(\mathbb{R})$,
defined by condition
$$
F(s) = S(s)\,F(-s)\,,
$$
where the reflection coefficient $S(s)$ is given by
$$
S(s) = \frac{M(s)}{M(-s)}\,.
$$
Incidentally the same reflection coefficient appears in the discussion of the zero modes in the Liouville model in~\cite{FV}.

It is evident that integral operator $P$ with kernel
$$
P(s,s^{\prime}) = \frac{1}{2}\left[\delta(s-s^{\prime})+S(s)\,\delta(s+s^{\prime})\right]
$$
defines a projection and $U$ maps $L_2(\mathbb{R})$ into subspace  $P\,L_2(\mathbb{R})$. However the natural completeness
$$
\int_{-\infty}^{+\infty} \mathrm{d}s\,\mathrm{d}s^{\prime}\,
P(s,s^{\prime})\,\overline{U(x,s)}\,U(y,s^{\prime}) = \delta(x-y)
$$
reduces to~(\ref{compl}) due to the fact, that $\phi(x,s)$ is an even function of $s$ and property $\overline{M(s)} = M(-s)$. The inversion $s\to -s$ is evidently connected to the Weyl reflection.
The proof of Kashaev results is given in Appendix.

After all we obtain the following expression for the eigenfunction of the undressed Casimir operator
\begin{equation}\label{Phip}
\Psi_p(x_1,x_2) = e^{-2\pi i p\, x_1}\,\frac{e^{-2\pi i s_1\,x_{21}}}{\gamma(x_{21}+p-s_2)}\,\phi(x_{21}-s_2, s_3)
\end{equation}
and now we can use them to formulate the orthogonality and
completeness for the kernel $S(x_1,x_2,x_3)$.

\section{Undressing of the eigenfunctions}

First of all we have to find out the connection between
$\Psi_p(x_1,x_2)$ and undressed eigenfunction
$\mathrm{A}^{-1}\,S(x_1,x_2,x_3)$.
The explicit expression for
the undressed eigenfunction reads
$$
\mathrm{A}^{-1}\,S(x_1,x_2,x_3) =
S_0\,c^{-1}\,
e^{-2\pi i \left(s_1\, x_{23}+s_2\, x_{31}+s_3\, x_{21}\right)}
\,\frac{1}{\gamma(x_{12}-s_1+2\,s_2)}\,
\frac{\gamma(x_{31}-\omega^{\prime\prime})}
{\gamma(x_{31}+s_1-s_2-s_3)}\,\cdot
$$
$$
\cdot\int_{\mathbb{R}}
\mathrm{d}\,t\,
e^{2\pi i \,t (s_2-s_1-s_3-\omega^{\prime\prime})}\,
\frac{\gamma(-t-\omega^{\prime\prime}+i\,0)
\gamma(t-x_{3}+s_3-s_2-\omega^{\prime\prime}+i\,0)}
{\gamma(x_{1}-t+s_2+s_3+\omega^{\prime\prime}+i\,0)\,\gamma(t-x_{3}-s_1)}\,,
$$
where undressing
operator $\mathrm{A}^{-1}$ is given by~(\ref{Akern}).
The $t$-integral is reduced to~(\ref{FT3eps}) and can be calculated
in explicit form so that we obtain
$$
\mathrm{A}^{-1}\,S(x_1,x_2,x_3) =
S_0\,e^{2\pi i \omega^{\prime\prime\,2}}\,
\frac{e^{-i\pi (s_2-s_1+s_3)^2}}
{\gamma(s_2-s_1-s_3)}
\,\cdot
$$
$$
\cdot
\frac{e^{2\pi i (s_1+s_3)\,x_{12}}}{\gamma(x_{12}+s_2+s_3+\omega^{\prime\prime})}
\,
\frac{e^{2\pi i (\omega^{\prime\prime}-s_3)\, x_{13}}\,\gamma(x_{23}+s_3-s_2-\omega^{\prime\prime})}
{\gamma(x_{13}+\omega^{\prime\prime})}\,.
$$
Next step is the calculation of Fourier transformation with respect variable $x_3$ using~(\ref{FT2eps})
$$
S_p(x_1,x_2) = \int_{\mathbb{R}}
\mathrm{d}\,x_3\, e^{-2\pi i \,p x_3}\,
\mathrm{A}^{-1}\,S(x_1,x_2,x_3)
=
$$
$$
=
S_0\,e^{2\pi i \omega^{\prime\prime\,2}}\,
\frac{e^{-i\pi (s_2-s_1+s_3)^2}}
{\gamma(s_2-s_1-s_3)}\,
\frac{e^{2\pi i (s_1+s_3)\,x_{12}}}
{\gamma(x_{12}+s_2+s_3+\omega^{\prime\prime})}
\,\cdot
$$
$$
\cdot\, e^{-2\pi i \, p\, x_1}\,
\int_{\mathbb{R}}\mathrm{d}\,t\, e^{-2\pi i \,
t\,(p-s_3+\omega^{\prime\prime})}\,
\frac{\gamma(x_{21}+s_3-s_2-\omega^{\prime\prime}+i\,0-t)}
{\gamma(\omega^{\prime\prime}-i\,0-t)} =
$$
$$
=
S_0\,e^{2\pi i \omega^{\prime\prime\,2}}\,
c\,
\frac{e^{-i\pi (s_2-s_1+s_3)^2}\,\gamma(p-s_3)}
{\gamma(s_2-s_1-s_3)}
\,\frac{e^{-2\pi i \, p\, x_{1}}}
{\gamma(x_{21}-s_2+p)}\,
\frac{e^{2\pi i (s_1+s_3)\,x_{12}}\,
\gamma(x_{21}+s_3-s_2-\omega^{\prime\prime})}
{\gamma(x_{12}+s_2+s_3+\omega^{\prime\prime})}\,.
$$
This expression coincides with~(\ref{Phip})
$$
S_p(x_1,x_2)
= \mathrm{Z}(s_1,s_2|s_3,p)\,\Psi_p(x_1,x_2)
$$
up to overall normalization $\mathrm{Z}(s_1,s_2|s_3,p)$
$$
\mathrm{Z}(s_1,s_2|s_3,p) = S_0\,i\,e^{2\pi i \omega^{\prime\prime\,2}}\,
\frac{e^{ -i\pi s_3^2 -2\pi i s_3(s_2+\omega^{\prime\prime})-i\pi (s_2-s_1+s_3)^2}\,\gamma(p-s_3)}
{\gamma(s_2-s_1-s_3)}\,.
$$
The special choice of the initial normalization constant $S_0$
\begin{equation}\label{norm}
S_0 = -i\,e^{-2\pi i \omega^{\prime\prime\,2}}\,
e^{i\pi s_3^2 +2\pi i s_3(s_2+\omega^{\prime\prime})+i\pi (s_2-s_1+s_3)^2}\,\gamma(s_2-s_1-s_3) = e^{2\pi i s_3\omega^{\prime\prime}}\,e^{i\phi}\,,
\end{equation}
where $\phi$ is real phase, leads to simplification
$$
\mathrm{Z}(s_1,s_2|s_3,p)  = \gamma(p-s_3)\,
$$
so that we obtain properly normalized eigenfunctions
\begin{equation}\label{Sp}
S_p(x_1,x_2)
= e^{2 \pi i\,s_3\, p}\,\gamma(p-s_3)\,\frac{e^{-2\pi i \, p\, x_{1}}}
{\gamma(x_{21}-s_2+p)}\,
e^{-2\pi i s_1\,x_{21}}\,\phi_{s_3}(x_{21}-s_2)\,.
\end{equation}
Now we shall prove the orthogonality and completeness
of the functions $S(x_1,x_2,x_3)$ in momentum representation
$$
S(x_1,x_2,p) = \int_{\mathbb{R}}
\mathrm{d}\,x_3\, e^{-2\pi i \,p x_3}\,S(x_1,x_2,x_3)\,.
$$
It will be sufficient to prove the orthogonality and completeness for the undressed eigenfunctions $S_p(x_1,x_2)$ because the dressed eigenfunction $S(x_1,x_2,p) = \mathrm{A}\,S_p(x_1,x_2)$ is obtained from the $S_p(x_1,x_2)$ after action of the dressing unitary operator $\mathrm{A}$.
Due to unitarity this dressing operator effectively cancels out from considered
relations.

Let us begin from orthogonality. There the dressing operator $\mathrm{A}$ cancels out on the first step
$$
\int_{\mathbb{R}^2} \mathrm{d}\,x_1\mathrm{d}\,x_2\,
\overline{S\left(x_1\,,x_2\,,q\right)}\,
S\left(x_1\,,x_2\,,p\right) = \int_{\mathbb{R}^2} \mathrm{d}\,x_1\mathrm{d}\,x_2\,
\overline{S_q\left(x_1\,,x_2\right)}\,
S_p\left(x_1\,,x_2\right) =
$$
$$
= \frac{\gamma(p-s)}
{\gamma(q-s^{\prime})}\,
\int_{\mathbb{R}^2} \mathrm{d}\,x_1\mathrm{d}\,x_2\,
e^{-2\pi i \, (p-q)\, x_{1}}\,
\frac{\gamma(x_{21}-s_2+q)}
{\gamma(x_{21}-s_2+p)}\,
\overline{\phi(x_{21}-s_2,s^{\prime})}\,\phi(x_{21}-s_2,s) =
$$
$$
= \delta(p-q)\,\frac{\gamma(p-s)}
{\gamma(p-s^{\prime})}\,
\int_{\mathbb{R}} \mathrm{d}\,x\,
\overline{\phi(x , s^{\prime})}\,\phi(x , s) =
$$
$$
= \rho^{-1}(s)\,\delta(p-q)\,
\left[\delta(s-s^{\prime})+\delta(s+s^{\prime})\,
\frac{\gamma(p-s)}{\gamma(p+s)}\right]\,.
$$
Note that the appearance of the second term containing    $\delta\left(s+s^{\prime}\right)$ and kernel of the intertwining
operator~(\ref{AI}) is the direct consequence of the equivalence of representations $\pi_s$ and $\pi_{-s}$.

The completeness for the undressed eigenfunction can be proven as follows
$$
\int_{0}^{+\infty} \mathrm{d}s\,\rho(s)\,\int_{\mathbb{R}} \mathrm{d}\,p\,
\overline{S_{p}\left(x_1^{\prime}\,,x_2^{\prime}\right)}\,
S_p\left(x_1\,,x_2\right) =
$$
$$
= \int_{\mathbb{R}} \mathrm{d}\,p\,e^{-2\pi i \, p\, (x_{1}-x_{1}^{\prime})}\,
\frac{\gamma(x_{21}^{\prime}-s_2+p)}
{\gamma(x_{21}-s_2+p)}\,
\int_{0}^{+\infty} \mathrm{d}s\,\rho(s)\,
\overline{\phi(x_{21}^{\prime}-s_2,s)}\,\phi(x_{21}-s_2,s) =
$$
$$
= \int_{\mathbb{R}} \mathrm{d}\,p\,e^{-2\pi i \, p\, (x_{1}-x_{1}^{\prime})}\,
\frac{\gamma(x_{21}^{\prime}-s_2+p)}
{\gamma(x_{21}-s_2+p)}\,
\delta(x_{21}^{\prime}-x_{21}) = \delta(x_{1}^{\prime}-x_{1})
\,\delta(x_{2}^{\prime}-x_{2})\,.
$$
Due to unitarity of dressing operator the same relation holds for the dressed eigenfunctions
$$
\int_{0}^{+\infty} \mathrm{d}s\,\rho(s)\,\int_{\mathbb{R}} \mathrm{d}\,p\,
\overline{S\left(x_1^{\prime}\,,x_2^{\prime}\,,p\right)}\,
S\left(x_1\,,x_2\,,p\right) = \delta(x_{1}^{\prime}-x_{1})
\,\delta(x_{2}^{\prime}-x_{2})\,.
$$
\vspace{1cm}

\section*{Acknowledgement}

One of the authors (L.D.F) is grateful to Professor
J.Teschner for discussions. In particular, his
unpublished notes~\cite{T0} received in the fall of 2012
contain procedure similar to our undressing
\footnote{Last week these notes were published as a part of~\cite{Tesch}.}.

Our work is partially supported by by RFBR grants
11-01-00570-a, 11-01-12037-ofi-m and the RAS
program "Mathematical problems of nonlinear dynamics".

\appendix
\renewcommand{\theequation}{\Alph{section}.\arabic{equation}}
\setcounter{table}{0}
\renewcommand{\thetable}{\Alph{table}}

%\section*{Appendices}

%%%%%%%%%%%%%%%%%%%%%%%%%%%%%%%%%%%%%%%%%%%%%%%%%%%%%%%%%%%%%%%%%%%%%%%%%%%%%%%%%%%%%%%%%%%%

%%%%%%%%%%%%%%%%%%%%%%%%%%%%%%%%%%%%%%%%%%%%%%

%%%%%%%%%%%%%%%%%%%%%%%%%%%%%%%%%%%%%%%%%%%%%%%%%%%%%%%%%%%%%%%%%%%%%%%%%%%%%%%%%%%%%%%%%%%%

\section{Orthogonality and completeness of Kashaev eigenfunctions}
\setcounter{equation}{0}

We shall need the generalization of identity~(\ref{FT2eps}) in the form
\begin{equation}\label{F1}
\int_{\mathbb{R}}
\mathrm{d}\,t\, e^{-2\pi i \,t\, s}
\,\frac{\gamma(t+a)}{\gamma(t+b)} =
c\,e^{2\pi i\,s(b-\omega^{\prime\prime})}\,
\frac{\gamma(a-b+\omega^{\prime\prime})
\,\gamma(-s-\omega^{\prime\prime})}
{\gamma(a-b-s+\omega^{\prime\prime})} =
\end{equation}
\begin{equation}\label{F2}
= c^{-1}\,
e^{2\pi i\,s(a+\omega^{\prime\prime})}
\,\frac{\gamma(b-a+s-\omega^{\prime\prime})}
{\gamma(b-a-\omega^{\prime\prime})
\,\gamma(s+\omega^{\prime\prime})}\,,
\end{equation}
where integral converges under conditions
\begin{equation}\label{cond}
\Im(s) < 0 \,\,,\,
\Im(a-b-s) < 0\,.
\end{equation}
The inverse formula is
\begin{equation}\label{invF}
\frac{\gamma(t+a)}{\gamma(t+b)} =
\frac{1}{c\,\gamma(b-a-\omega^{\prime\prime})}\,
\int_{\mathbb{R}}
\mathrm{d}\,s\, e^{2\pi i \,s\,(t+a+\omega^{\prime\prime})}
\,\frac{\gamma(s+b-a-\omega^{\prime\prime})}
{\gamma(s+\omega^{\prime\prime})}\,,
\end{equation}
and the contour goes below the singularity at $s = 0$.

\subsection{Orthogonality}

We take eigenfunction in the form
$$
\phi(x,\lambda) = e^{-i\pi(x-\omega^{\prime\prime})^2}\,
\gamma(x+\lambda-\omega^{\prime\prime})\,
\gamma(x-\lambda-\omega^{\prime\prime})\,,
$$
so that
$$
\overline{\phi(x,\lambda)}\,\phi(x,\mu) =
e^{4\pi i \,x\,\omega^{\prime\prime}}\,
\frac{\gamma(x+\mu-\omega^{\prime\prime})\,
\gamma(x-\mu-\omega^{\prime\prime})}
{\gamma(x+\lambda+\omega^{\prime\prime})\,
\gamma(x-\lambda+\omega^{\prime\prime})}\,,
$$
and the singularities here have to be understood as
$
\omega^{\prime\prime} \to \omega^{\prime\prime} - i\,0\,.
$
We have to calculate
$$
I(\lambda,\mu) = \int_{\mathbb{R}} \mathrm{d}\,x\,
\overline{\phi(x,\lambda)}\,\phi(x,\mu)\,.
$$
This function is even in $\lambda$ and $\mu$ and it is sufficient to calculate it in one quadrant, say $\lambda > 0$ and $\mu < 0$.
First we transform the ratio of two $\gamma$-functions using~(\ref{invF})
$$
\frac{\gamma(x+\mu-\omega^{\prime\prime})}
{\gamma(x+\lambda+\omega^{\prime\prime})} =
\frac{1}{c\,\gamma(\lambda-\mu+\omega^{\prime\prime})}\,
\int_{\mathbb{R}}
\mathrm{d}\,s\, e^{2\pi i \,s\,(x+\mu)}
\,\frac{\gamma(s+\lambda-\mu+\omega^{\prime\prime})}
{\gamma(s+\omega^{\prime\prime})}\,,
$$
calculate $x$-integral using~(\ref{F2})
$$
\int_{\mathbb{R}}
\mathrm{d}\,x\, e^{2\pi i \,x\,(s+2\omega^{\prime\prime})}
\,\frac{\gamma(x-\mu-\omega^{\prime\prime})}
{\gamma(x-\lambda+\omega^{\prime\prime})} =
c^{-1}\,e^{2\pi i\,\mu(s+2\omega^{\prime\prime})}
\,\frac{\gamma(\mu-\lambda-s-\omega^{\prime\prime})}
{\gamma(\mu-\lambda+\omega^{\prime\prime})
\,\gamma(-s-\omega^{\prime\prime})}\,,
$$
and after these two steps arrive to the following expression for $I(\lambda,\mu)$
$$
I(\lambda,\mu) =
\frac{1}{c^2\,\gamma(\lambda-\mu+\omega^{\prime\prime})\,
\gamma(\mu-\lambda+\omega^{\prime\prime})}\,
\int_{\mathbb{R}}
\mathrm{d}\,s\, e^{4\pi i \,s\,\mu}
\,\frac{\gamma(s+\lambda-\mu+\omega^{\prime\prime})\,
\gamma(\mu-s-\lambda-\omega^{\prime\prime})}
{\gamma(s+\omega^{\prime\prime})\,
\gamma(-s-\omega^{\prime\prime})}\,.
$$
The ratio of $\gamma$-functions is reduced to the simple exponent
$$
\frac{\gamma(s+\lambda-\mu+\omega^{\prime\prime})\,
\gamma(\mu-s-\lambda-\omega^{\prime\prime})}
{\gamma(s+\omega^{\prime\prime})\,
\gamma(-s-\omega^{\prime\prime})} =
e^{i\pi(s+\lambda-\mu+\omega^{\prime\prime})^2 -
i\pi(s+\omega^{\prime\prime})^2}
$$
due to reflection relation~(\ref{reflex}) so that we obtain for $\lambda > 0$ and $\mu < 0$
$$
I(\lambda,\mu) =
\frac{e^{i\pi(\lambda-\mu)^2 + 2i\pi\omega^{\prime\prime}(\lambda-\mu)}}
{c^2\,\gamma(\lambda-\mu+\omega^{\prime\prime})\,
\gamma(\mu-\lambda+\omega^{\prime\prime})}\,
\int
\mathrm{d}\,s\,
e^{2i\pi s(\lambda+\mu)} =
\frac{e^{4i\pi\lambda^2 + 4i\pi\lambda\omega^{\prime\prime}}}
{c^2\,\gamma(2\lambda+\omega^{\prime\prime})\,
\gamma(-2\lambda+\omega^{\prime\prime})}\,
\delta(\lambda+\mu)\,.
$$
The full answer is restored by the symmetry
$$
\int_{\mathbb{R}} \mathrm{d}\,x\,
\overline{\phi(x,\lambda)}\,\phi(x,\mu) =
\frac{1}{M(\lambda)\,M(-\lambda)}\,
\left[\,\delta(\lambda-\mu)+\delta(\lambda+\mu)\,\right]\,.
$$
where
$$
M(\lambda) = c\,e^{-2i\pi\lambda^2 - 2i\pi\lambda\omega^{\prime\prime}}\,
\gamma(2\lambda+\omega^{\prime\prime})\,.
$$
It is exactly the formula~(\ref{ort}) with
$$
\rho(\lambda) = M(\lambda)\,M(-\lambda) = e^{-4\pi i \,\lambda\,\omega^{\prime\prime}}\,
\frac{\gamma(2\lambda+\omega^{\prime\prime})}
{\gamma(2\lambda-\omega^{\prime\prime})} =
\left(e^{\frac{i\pi \lambda}{\omega}} - e^{-\frac{i\pi \lambda}{\omega}}\right)\left(e^{\frac{i\pi \lambda}{\omega^\prime}} -
e^{-\frac{i\pi \lambda}{\omega^\prime}}\right)\,.
$$

\subsection{Completeness}

Now we have to calculate
$$
I(x,y) = \int_{0}^{+\infty} \mathrm{d}\lambda\,\rho(\lambda)\,
\overline{\phi(x,\lambda)}\,\phi(y,\lambda) = \int_{-\infty}^{+\infty} \mathrm{d}\lambda\,\sigma(\lambda)\,
\overline{\phi(x,\lambda)}\,\phi(y,\lambda)\,,
$$
where we used the symmetry $\lambda\to -\lambda$
$$
\rho(\lambda) = \sigma(\lambda)+\sigma(-\lambda)\ \ ;\ \
\sigma(\lambda) = e^{-4\pi i \lambda\,\omega^{\prime\prime}}-
e^{-4\pi i \lambda\,\left(\omega^{\prime\prime}-2\omega^{\prime}\right)}\,.
$$
We introduce regularization and obtain the following expression
$$
I(x,y) = e^{i\pi(x^2-y^2)-2i\pi(x-y)\omega^{\prime\prime}}\,
\int_{\mathbb{R}}
\mathrm{d}\lambda\,\sigma(\lambda)\,e^{2\pi \lambda \delta}
\frac{\gamma(y+\lambda-\omega^{\prime\prime}+i\epsilon)\,
\gamma(y-\lambda-\omega^{\prime\prime}+i\epsilon)}
{\gamma(x+\lambda+\omega^{\prime\prime}-i\epsilon)
\,\gamma(x-\lambda+\omega^{\prime\prime}-i\epsilon)}\,,
$$
where $\epsilon > 0 \,, \delta > 0$ and $\delta > 2\epsilon$.
First we transform the ratio of $\gamma$-functions using~(\ref{invF})
$$
\frac{\gamma(y+\lambda-\omega^{\prime\prime}+i\epsilon)}
{\gamma(x+\lambda+\omega^{\prime\prime}-i\epsilon)} =
\frac{1}{c\,\gamma(x-y+\omega^{\prime\prime}-2i\epsilon)}\,
\int_{\mathbb{R}}
\mathrm{d}\,s\, e^{2\pi i \,s\,(y+\lambda+i\epsilon)}
\,\frac{\gamma(s+x-y+\omega^{\prime\prime}-2i\epsilon)}
{\gamma(s+\omega^{\prime\prime})}\,,
$$
where the contour goes below the singularity at $s = 0$.
Let us consider the $\lambda$-integral with the first
contribution in $\sigma(\lambda)$ and make change of variables $\lambda\to -\lambda$ for convenience
$$
I_1(s) =
\int
\mathrm{d}\,\lambda\, e^{-2\pi i \,\lambda\,
(s-2\omega^{\prime\prime}+i\delta)}
\,
\frac{\gamma(y+\lambda-\omega^{\prime\prime}+i\epsilon)}
{\gamma(x+\lambda+\omega^{\prime\prime}-i\epsilon)}
\,.
$$
The second condition in~(\ref{cond}) is fulfilled due to relation $\delta > 2\epsilon$ and using~(\ref{F2}) we obtain
$$
I_1(s) =
c^{-1}\,e^{2\pi i\,(y+i\epsilon)(s-2\omega^{\prime\prime}+i\delta)}
\,\frac{\gamma(x-y+s-\omega^{\prime\prime}+i(\delta-2\epsilon))}
{\gamma(x-y+\omega^{\prime\prime}-2i\epsilon)
\,\gamma(s-\omega^{\prime\prime}+i\delta)}\,.
$$
The same calculation with the second contribution in $\sigma(\lambda)$ gives
$$
I_2(s) =
\int
\mathrm{d}\,\lambda\, e^{-2\pi i \,\lambda\,
(s-2\omega+2\omega^{\prime}+i\delta)}
\,
\frac{\gamma(y+\lambda-\omega^{\prime\prime}+i\epsilon)}
{\gamma(x+\lambda+\omega^{\prime\prime}-i\epsilon)} =
\,
$$
$$
=
c^{-1}\,e^{2\pi i\,(y+i\epsilon)(s-2\omega+2\omega^{\prime}+i\delta)}
\,\frac{\gamma(x-y+s+\omega^{\prime\prime}-2\omega+2\omega^{\prime}+i(\delta-2\epsilon))}
{\gamma(x-y+\omega^{\prime\prime}-2i\epsilon)
\,\gamma(s+\omega^{\prime\prime}-2\omega+2\omega^{\prime}+i\delta)}\,.
$$
The change of variables $s \to s-2\omega^{\prime}$ in $s$-integral
containing $I_2(s)$ transforms ratio of $s$-dependent $\gamma$-functions to the form
$$
\frac{\gamma(x-y+s+\omega-\omega^{\prime}-2i\epsilon)}
{\gamma(s+\omega-\omega^{\prime})}
\,\frac{\gamma(x-y+s-\omega+\omega^{\prime}+i(\delta-2\epsilon))}
{\gamma(s-\omega+\omega^{\prime}+i\delta)}
$$
where we restored $s$-dependent $\gamma$-functions
from the first stage.
In the corresponding $s$-integral containing $I_1(s)$ there is the following
ratio of $s$-dependent $\gamma$-functions
$$
\frac{\gamma(x-y+s+\omega^{\prime\prime}-2i\epsilon)}
{\gamma(s+\omega^{\prime\prime})}
\,\frac{\gamma(x-y+s-\omega^{\prime\prime}+i(\delta-2\epsilon))}
{\gamma(s-\omega^{\prime\prime}+i\delta)}\,.
$$
The formula
$$
\gamma(z+\omega-\omega^{\prime})\,
\gamma(z-\omega+\omega^{\prime}) = \gamma(z+\omega^{\prime\prime})\,\gamma(z-\omega^{\prime\prime})
$$
allows to transform one expression to another for $\delta = 0$.
It means that in situation when it is possible to
skip $\delta$-regularization, we obtain the
integral over closed contour.
Let us deform the contour in integral
with $I_1(s)$: the contour which goes above the singularity at $s = 0$ and
a small closed contour around $s=0$ leading to additional contribution
$$
2\pi i\, \mathrm{Res}_{s=0} = \frac{1}{c}\frac{\gamma(x-y+\omega^{\prime\prime}-2i\epsilon)
\,\gamma(x-y-\omega^{\prime\prime}-2i\epsilon+i\delta)}
{\gamma(-\omega^{\prime\prime}+i\delta)}\,.
$$
In remaining two integrals it is possible to put $\delta=0$ and therefore to reduce it to the integral over closed contour without any singularity inside.
As a result only the term $2\pi i\, \mathrm{Res}_{s=0}$ leads to nonzero contribution and restoring all needed factors we obtain
$$
\frac{c\,\gamma(x-y-\omega^{\prime\prime}-2i\epsilon+i\delta)}
{\gamma(x-y+\omega^{\prime\prime}-2i\epsilon)
\gamma(-\omega^{\prime\prime}+i\delta)}
\to \frac{1}{2\pi i}\,
\frac{i\delta}
{(x-y-2i\epsilon)(x-y+i\delta-2i\epsilon)}
 \to \delta(x-y)\,,
$$
so that
$$
\int_{0}^{+\infty} \mathrm{d}\lambda\,\rho(\lambda)\,
\overline{\phi(x,\lambda)}\,\phi(y,\lambda) = \delta(x-y)\,.
$$


\vspace{1cm}

\begin{thebibliography}{99}

\bibitem{FT}
L.D. Faddeev and L.A. Takhtajan,	
{\em Liouville model on the lattice},
Lect.Notes Phys. 246 (1986) 166-179

\bibitem{G}
J.-L. Gervais,
{\em Solving the strongly coupled 2d gravity: 1. Unitary truncation and
quantum group structure},
Commun. Math. Phys. 138, 301–338 (1991)

\bibitem{ZZ}
Alexander B. Zamolodchikov and Alexei B. Zamolodchikov, 	
{\em Structure constants and conformal bootstrap in Liouville field theory},
Nucl.Phys. B477 (1996) 577-605
e-Print: hep-th/9506136

\bibitem{F} L. D. Faddeev,
{\em Modular double of a  quantum group}, Conf. Mosh\'e Flato 1999,
vol. I, Math. Phys. Stud. {\bf 21}, Kluwer, Dordrecht, 2000, pp. 149--156.
arXiv:math/9912078.

\bibitem{BT}
A.G. Bytsko and J. Teschner,	
{\em R operator, coproduct and Haar measure for the modular double of $U_q(sl(2,R))$},
Commun.Math.Phys. 240 (2003) 171-196
e-Print: math/0208191

\bibitem{PT}
B. Ponsot and J. Teschner,	
{\em Clebsch-Gordan and Racah-Wigner coefficients for a continuous series of representations of $U_q(sl(2,R))$},
Commun.Math.Phys. 224 (2001) 613-655
e-Print: math/0007097

\bibitem{T}
J. Teschner,	
{\em Liouville theory revisited},
Class.Quant.Grav. 18 (2001) R153-R222
e-Print: hep-th/0104158

\bibitem{D}
V. G. Drinfeld,
{\em Quantum groups},
J. Sov. Math. 41 (1988) 898
[Zap.Nauchn. Semin. 155 (1986) 18].

\bibitem{J}
M.Jimbo,
{\em A q-differnece analogue of U(g) and the Yang-Baxter equation},
Lett.Math.Phys.10(1985) 63-69


\bibitem{K}
R. M. Kashaev,
{\em On the spectrum of Dehn twists in quantum Teichmuller theory},
in Physics and combinatorics (Nagoya, 2000), pages 63–81, World Sci. Publ., River Edge, NJ, 2001,
e-Print: math/0008148

\bibitem{K1}
R.M. Kashaev, {\em Discrete Liouville equation and Teichmuller theory}, arXiv:0810.4352, to
appear in: Handbook in Teichmuller theory, Vol. III, IRMA Lect. Math. Theor. Phys.

\bibitem{FKV}	
L.D. Faddeev, R.M. Kashaev and  A.Yu. Volkov,
{\em Strongly coupled quantum discrete Liouville theory.
1. Algebraic approach and duality}
Commun.Math.Phys. 219 (2001) 199-219,
e-Print: hep-th/0006156

\bibitem{V}
A. Yu. Volkov,
{\em Noncommutative Hypergeometry},
Commun. Math.Phys. 258 (2005) 257,
arXiv:math/0312084

\bibitem{FV}
L.D. Faddeev and A.Yu. Volkov,
{\em Discrete evolution for the zero-modes of the Quantum Liouville Model.}
J.Phys. A41 (2008) 194008
e-Print: arXiv:0803.0230

\bibitem{T0}
J. Teschner,	
{\em The Clebsch-Gordan maps for the modular double},
unpublished notes

\bibitem{Tesch}
I. Nidaiev, J. Teschner,	
{\em On the relation between the modular double
of $U_q(sl(2,R))$ and the quantum Teichmueller theory}
e-Print: arXiv:1302.3454

\end{thebibliography}
\end{document}